\documentclass[english,twocolumn,showpacs,prl,preprintnumbers,amsmath,amssymb,superscriptaddress]{revtex4}
\usepackage{graphicx}

\makeatletter

\makeatletter


\makeatother

\makeatother

\usepackage{babel}
\makeatother
\begin{document}

\preprint{XXX}

\title{Bright Source of Cold Ions for Surface-Electrode Traps}

\author{Marko Cetina}

\affiliation{Department of Physics, MIT-Harvard Center for Ultracold Atoms, and
Research Laboratory of Electronics, Massachusetts Institute of Technology,
Cambridge, Massachusetts 02139, USA}

\author{Andrew Grier}

\affiliation{Department of Physics, MIT-Harvard Center for Ultracold Atoms, and
Research Laboratory of Electronics, Massachusetts Institute of Technology,
Cambridge, Massachusetts 02139, USA}

\author{Jonathan Campbell}

\affiliation{Department of Physics, United States Military Academy, West Point,
NY, 10996}

\author{Isaac Chuang}

\affiliation{Department of Physics, MIT-Harvard Center for Ultracold Atoms, and
Research Laboratory of Electronics, Massachusetts Institute of Technology,
Cambridge, Massachusetts 02139, USA}

\author{Vladan Vuleti\'{c}}

\affiliation{Department of Physics, MIT-Harvard Center for Ultracold Atoms, and
Research Laboratory of Electronics, Massachusetts Institute of Technology,
Cambridge, Massachusetts 02139, USA}

\date{\today}

\begin{abstract}
We produce large numbers of low-energy ions by photoionization of
laser-cooled atoms inside a surface-electrode-based Paul trap. The
isotope-selective trap loading rate of $4\times10^{5}$ Yb$^{+}$
ions/s exceeds that attained by photoionization (electron impact ionization)
of an atomic beam by four (six) orders of magnitude. Traps as shallow
as 0.13 eV are easily loaded with this technique. The ions are confined
in the same spatial region as the laser-cooled atoms, which will allow
the experimental investigation of interactions between cold ions and
cold atoms or Bose-Einstein condensates. 
\end{abstract}

\pacs{42.50.Dv, 03.67.Hk, 42.50Gy, 32.80.Pj}

\maketitle

Among many candidate systems for large-scale quantum information processing,
trapped ions currently offer unmatched coherence and control properties
\cite{Zoller05}. The basic building blocks of a processor, such as
quantum gates \cite{Monroe95}, subspaces with reduced decoherence
\cite{Kielpinski01}, quantum teleportation \cite{Barrett04,Riebe04},
and entanglement of up to eight ions \cite{Leibfried05,Haffner05}
have already been demonstrated. Nevertheless, since a logical qubit
will likely have to be encoded simultaneously in several ions for
error correction \cite{Shor95,Steane96}, even a few-qubit system
will require substantially more complex trap structures than currently
in use. Versatile trapping geometries can be realized with surface-electrode
Paul traps, where electrodes residing on a surface create three-dimensional
confining potentials above it \cite{Chiaverini05}. In contrast to
three-dimensional traps \cite{Stick06,Rowe02,Deslauriers06}, such
surface traps can be patterened using standard lithographic techniques,
and allow increased optical access to the ions and real-time control
over their position in all directions.

While the prospect of scalable quantum computing has been the main
motivation for developing surface-electrode traps, it is likely that
this emerging technology will have a number of other important, and
perhaps more immediate, applications. Porras and Cirac have proposed
using dense lattices of ion traps, where neighboring ions interact
via the Coulomb force, for quantum simulation \cite{Porras04}. A
lattice, with a larger period to avoid ion-ion interactions altogether,
could allow the parallel operation of many single-ion optical clocks
\cite{Oskay06}, thereby significantly boosting the signal-to-noise
ratio. The increased optical access provided by planar traps could
be used to couple a linear array of ion traps to an optical resonator
and efficiently map the stored quantum information onto photons \cite{Keller04}.
Since the surface-electrode arrangement allows one to move the trap
minimum freely in all directions, ions can be easily embedded in an
ensemble of cold neutral atoms for investigations of cold ion-atom
collisions \cite{Cote2000_IonAtomCollisions}, charge transport \cite{Cote2000_ChargeHopping},
or even the interaction of a single ion with a Bose-Einstein condensate
\cite{Cote2002_IonBEC}.

Compared to standard Paul traps \cite{Stick06,Rowe02,Deslauriers06},
the open geometry of surface-electrode traps restricts the trap depth
and increases the susceptibility to stray electric fields, making
trap loading and compensation more difficult. Nevertheless, successful
loading from a thermal atomic beam has recently been demonstrated
using photoionization \cite{Seidelin06} or electron-impact ionization
aided by buffer gas cooling \cite{Brown07pra}. The former \cite{Kjaergaard00,Gulde01}
is superior in that it provides faster, isotope-selective loading
\cite{Wunderlich06,Seidelin06,Lucas04,Keller03}. However, the loading
rate and efficiency remain relatively low, and charge exchange collisions
make it difficult to load pure samples of rare isotopes \cite{Lucas04}.

\begin{figure}
\includegraphics[clip]{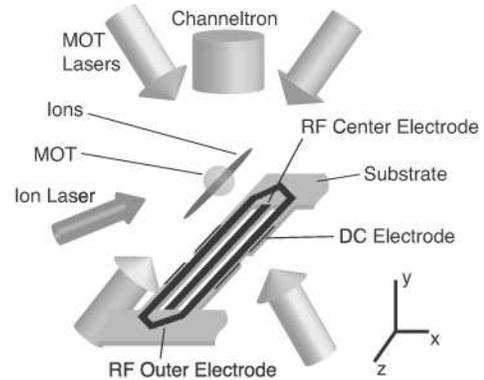}

\caption{Setup for loading a surface-electrode ion trap by in-trap photoionization
of laser-cooled atoms. A $^{172}$Yb or $^{174}$Yb magneto-optical
trap (MOT) is formed 4 mm above the trap surface. Cold Yb$^{+}$ ions
are produced inside the Paul trap by single-photon excitation from
the excited $^{1}P_{1}$ atomic state. \label{chip} }
\end{figure}

In this Letter, we demonstrate that large numbers of low-energy ions
can be produced by photoionization of a laser-cooled, isotopically
pure atomic sample, providing a robust and virtually fail-safe technique
to load shallow or initially poorly compensated surface ion traps.
We achieve a loading rate of $4\times10^{5}$ Yb ions per second into
a $U_{0}=0.4$ eV deep printed-circuit ion trap, several orders of
magnitude larger than with any other method demonstrated so far, and
have directly loaded traps as shallow as $U_{0}=0.13$ eV. The trapping
efficiency for the generated low-energy ions is of order unity. We
also realize the first system where ions are confined in the same
spatial region as laser-cooled atoms, allowing for future experimental
studies of cold ion-atom collisions.

Efficient photoionization of Yb atoms is accomplished with a single
photon from the excited $^{1}P_{1}$ state that is populated during
laser cooling, and that lies 3.11 eV, corresponding to a 394 nm photon,
below the ionization continuum. Due to momentum and energy conservation,
most of the ionization photon's excess energy is transferred to the
electron. Therefore, when we ionize atoms at rest even with 3.36 eV
(369 nm) photons -- the ion cooling light -- the calculated kinetic
energy of the ions amounts to only 8 mK (0.7 $\mu$eV). Every ion
generated inside the trap should therefore be captured, and we easily
observe ion trapping even without subsequent laser cooling.

\begin{figure}
\includegraphics[clip,width=3in,keepaspectratio]{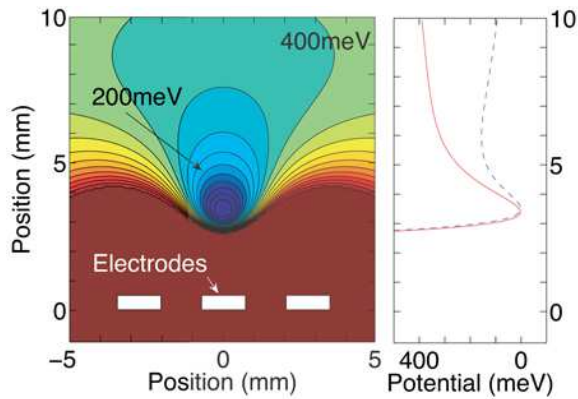}

\caption{Left: Trapping pseudopotential for RF amplitudes of $V_{o}=560$
V and $V_{c}=-340$ V applied to the outer and center electrodes together
with a DC bias of -1.25 V applied to all electrodes. The contours
are spaced by 50 meV. Right: The dc-unbiased trap (dashed blue curve)
exhibits less micromotion heating but is significantly shallower than
the dc-biased trap (solid red curve). (Color Online)}

\label{trapping_potential} 
\end{figure}

The ion trap is a commercial printed circuit on a vacuum-compatible
substrate (Rogers 4350) with low radiofrequency (RF) loss. The three
1 mm-wide, 17.5 $\mu$m-thick copper RF electrodes are spaced by 1
mm wide slits (Fig.\ref{chip}), whose inner surfaces are metallized
to avoid charge buildup on dielectric surfaces. The two outer RF electrodes
are electrically connected. Twelve dc electrodes placed outside the
RF electrodes provide trapping in the axial direction, and permit
cancellation of stray electric fields. In addition, the RF electrodes
can be dc-biased to apply a vertical electric field. All dielectric
surfaces outside the dc electrodes have been removed with the exception
of a 500 $\mu$m strip supporting the dc electrodes.

The ratio between the RF voltages $V_{c}$ (applied to the center
electrode) and $V_{o}$ (outer electrode) determines the trap height
above the surface. With a typical value of $V_{c}/V_{o}=-0.63$, the
RF node is located 3.6(1) mm above from the surface. For $V_{o}=540$
V applied to the outer electrodes, at an RF frequency of 850 kHz the
secular trap potential has a predicted depth of $U_{0}=0.16$ eV (Fig.\ref{trapping_potential})
and a measured secular frequency of 60 kHz . The trap can be deepened
by applying a static negative bias voltage $V_{dc}$ to all RF electrodes
\cite{Brown07pra}, Fig.\ref{trapping_potential}, and unbiased once
the ions are loaded. Using $V_{dc}=0.5$ V we were able to load traps
at RF voltages as low as $V_{o}=250$ V, corresponding to $U_{0}$=0.13
eV.

All Yb and Yb$^{+}$ cooling, detection and photoionization light
is derived from near-UV external-cavity diode lasers. $^{172}$Yb
or $^{174}$Yb atoms are laser-cooled in a magneto-optical trap (MOT)
using the $^{1}S_{0}\rightarrow{^{1}P_{1}}$ transition at 399 nm
\cite{Park03}. A master-slave laser system consisting of an external-cavity
grating laser and an injected slave laser using violet laser diodes
(Nichia Corp. NDHV310ACAE1) delivers 10 mW in three pairs of 1.7 mm
beams. The MOT, located 4 mm above the substrate, is loaded from an
atomic beam produced by a resistively heated oven placed 8 cm from
the trapping region. Typically $5\times10^{5}$ Yb atoms are loaded
into the MOT with a lifetime of 300 ms at an estimated temperature
of a few mK.

\begin{figure}
\includegraphics[clip,width=3in,keepaspectratio]{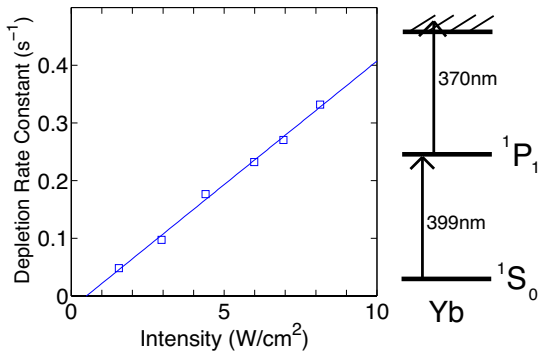}

\caption{Linear dependence of the loss rate constant of a $^{174}$Yb MOT
(rms width of 300 $\mu$m) on 370 nm beam peak intensity ($w=43$
$\mu$m), indicating that photoionization is accomplished with a single
370 nm photon from the $^{1}P_{1}$ state that is populated during
laser cooling with 399 nm light. }

\label{MOT_loss} 
\end{figure}

Photoionization from the excited $^{1}P_{1}$ state populated during
laser cooling is accomplished using either the ion cooling laser at
370 nm with a power of 750 $\mu$W and intensity of 850 mW/cm$^{2}$,
or a focused UV light-emitting diode (UV LED, Nichia Corp. NCCU001,
emission at (385$\pm$10) nm) with a power of 8.7 mW and intensity
of 125 mW/cm$^{2}$ at the MOT position. The efficiency of ionization
is manifest as a 30\% decrease in MOT atom number due to an increase
in the MOT decay rate constant by $\Gamma=0.3$ s$^{-1}$. The generated
ions are also detected directly with a Burle Magnum 5901 Channeltron
avalanche detector located 4 cm above the MOT. The UV-light induced
MOT loss depends linearly on UV laser intensity (Fig.\ref{MOT_loss}),
indicating that the ionization process involves a single 370 nm photon.
We have also confirmed that the dominant ionization proceeds from
the $^{1}P_{1}$ state: when we apply on/off modulation to both the
399 nm MOT light and the 370 nm UV light out of phase, such that the
UV light interacts only with ground-state atoms, we observe more than
13-fold decrease in ionization compared to in-phase modulation. From
the observed loss rate in combination with an estimated saturation
$s=0.6-2$ of the $^{1}S_{0}\rightarrow$$^{1}P_{1}$ MOT transition,
we determine a cross section $\sigma=4\times10^{-18}$cm$^{2}$ for
ionization of $^{174}$Yb with 370 nm light from the excited $^{1}P_{1}$
state. We estimate that this value is accurate to a factor of two,
due to uncertainties in the $^{1}P_{1}$ population.

The photoionization typically produces $4\times10^{5}$ cold ions
per second near the minimum of the pseudopotential. Trapped-ion detection
with the Channeltron provides rapid readout, making it particularly
useful for observing fast trap loading or searching for an initial
signal with a poorly compensated trap. Since the detector electric
field overwhelms the pseudopotential, we turn on the Channeltron in
1 $\mu$s using a Pockels cell driver which is fast compared to the
3.3 $\mu$s flight time of the ions. The Channeltron signal is calibrated
against fluorescence from a known number of ions, as described below.

\begin{figure}
\includegraphics[clip,width=3in,keepaspectratio]{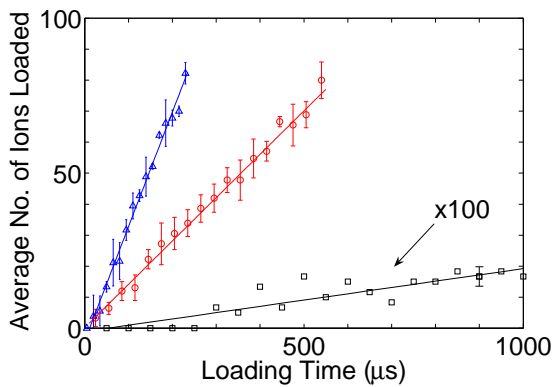}

\caption{Number of trapped $^{172}$Yb ions for photoionization loading with
the UV LED from an atomic beam (black squares), from a MOT (red circles),
and from a MOT with additional 370 nm laser light (blue triangles).
The fitted rates are 2.0$\times$$10^{2}$, 1.4$\times$$10^{5}$,
and 3.8$\times$$10^{5}$ ions/s, respectively. (Color online)}

\label{loading_rate} 
\end{figure}

We measure ion loading rates by varying the time between turning on
the trap and switching on the detector, which empties the trap. Given
the brightness of our cold-ion source, ion trapping is easily accomplished
even without laser cooling of the ions. Fig.\ref{loading_rate} shows
the loading rate for photoionization of atoms from the MOT and from
the atomic beam for the trap potential of depth of $U_{0}$=0.4 eV
shown in Figure\ref{trapping_potential}. The \emph{loading} rate
from the MOT, $4\times10^{5}$ ions/s, is three orders of magnitude
higher than that from the beam alone, compared to a ratio of only
four in the ion \emph{production} rates measured without trap. Thus
ions that were produced from laser-cooled atoms in a magneto-optical
trap are 200 times more likely to be trapped than ions produced from
the atomic beam. We ascribe this difference to the much higher MOT
atomic density near the ion-trap minimum and to the lower energy of
the produced ions. As the MOT is isotopically pure, our observations
of a $10^{3}$ loading rate ratio between MOT and atomic beam imply
an additional achievable factor of $10^{3}$ in isotope selectivity
beyond the 370:1 isotope selectivity resulting from spectrally selective
photoionization in an atomic beam \cite{Lucas04}.

From comparisons of electron-impact ionization loading and atomic-beam
photoionization loading performed by other groups \cite{Wunderlich06,Lucas04},
we conclude that our loading rate is six to seven orders of magnitude
higher than that achieved with traditional electron-impact ionization
and four orders of magnitude higher than all previous results. In
addition, by comparing the typical observed loss rate from the MOT
($1.1\times10^{5}$ atoms/s) to the typical observed loading rate
($2.4\times10^{5}$ ions/s) under similar conditions, we conclude
that our trapping efficiency is comparable to unity. We attribute
the discrepancy in rates to calibration of the Channeltron ion detector
and uncertainty in the MOT population. This efficiency may prove an
important advantage for suppressing anomalous ion heating that has
been linked to electrode exposure to the atomic beam during the loading
process \cite{Deslauriers06}. The large loading rate will also be
beneficial for applications that require a large, isotopically pure
sample, such as quantum simulation in an ion lattice \cite{Porras04}.

\begin{figure}
\includegraphics[width=3in,keepaspectratio]{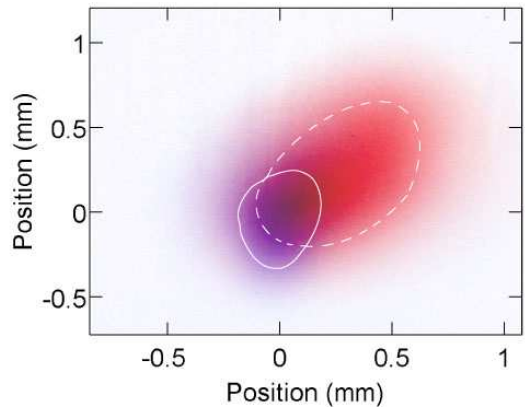}

\caption{Spatially overlapping ion and atom clouds. The false color image
shows a trapped $^{172}$Yb$^{+}$ ensemble containing $10^{2}$-$10^{3}$
ions (blue cloud lower left), placed inside a magneto-optical trap
containing $5\times10^{5}$ neutral $^{172}$Yb atoms (red cloud upper
right). The solid and dashed line indicate the half-maximum contour
for the ions and atoms, respectively. (Color online)}

\label{ion_MOT} 
\end{figure}

The loaded ions are cooled and observed via fluorescence using an
external-cavity grating laser. \cite{Kielpinski05}. To reach the
target wavelength of 369.525 nm with a 372 nm laser diode (Nichia)
in a Littrow setup with a first-order grating reflectivity of 28\%,
we cool the diode to temperatures between -10 and -20$^{\circ}$C
in a moisture-tight container. The laser provides an output power
of 1.5 mW and is continuously tunable over more than 10 GHz.

For laser cooling of Yb$^{+}$, the laser is typically tuned 200 MHz
below the $^{2}S_{1/2}\rightarrow{^{2}P_{1/2}}$ transition. An external-cavity
repumper laser operating at 935 nm is also necessary to empty the
long-lived $^{2}D_{3/2}$ state on the $^{2}D_{3/2}\rightarrow{^{3}D[3/2]_{1/2}}$
transition \cite{Bell91}. The UV light scattered by the ions is collected
with an aspheric lens of numerical aperture 0.40 placed inside the
vacuum chamber at a distance of 19 mm from the trap. The collected
light passes through an interference filter, and is evenly split between
a charge-coupled device camera and a photomultiplier. The maximum
photomultiplier count rate is 5000 s$^{-1}$ per ion. To calibrate
the Channeltron detector, we first cool a small cloud of ions (approximately
100) to below the crystallization temperature, as identified by a
sudden change in the fluorescence \cite{Diedrich87}. By comparing
the resonance fluorescence of the crystal to that of a single trapped
ion, we determine the absolute number of ions loaded, and subsequently
measure the Channeltron signal for the same cloud.

The optical observation of the trapped ions allows us to determine
the trap position and optimize the overlap with the MOT. We move the
MOT laterally using magnetic bias coils and the ion trap vertically
by changing the amplitude ratio of the voltages applied to the two
RF electrodes. Fig.\ref{ion_MOT} shows that for the optimal loading
position of the trap, the pseudopotential minimum is located inside
the neutral-atom cloud. We have thus demonstrated the first trapping
of cold ions and neutral atoms in the same spatial region, which will
allow the experimental investigation of cold ion-atom collisions.

In conclusion, we have realized a novel, simple and robust system
to load large numbers of low-energy ions into a versatile planar ion
trap. Further trap miniaturization while maintaining a large loading
rate can be achieved with a nested electrode design, where large outer
electrodes provide initial trapping, and a series of smaller inner
electrodes provides stronger confinement as the ions are transported
towards the surface. This system provides the means to realize ion
lattices for quantum simulation \cite{Porras04}, ionic quantum memory
with optical readout \cite{Keller04}, many-ion optical clocks, or
mixed ion-atom systems for the investigation of cold collisions \cite{Cote2000_IonAtomCollisions}
and charge transport \cite{Cote2000_ChargeHopping}.

We would like to thank Brendan Shields and Ken Brown for assistance.
This work was supported in part by the NSF Center for Ultracold Atoms.

\bibliographystyle{apsrev} \bibliographystyle{apsrev} 
\end{document}